n\documentclass[twocolumn,prl]{revtex4}%
\documentclass[twocolumn,prl]{revtex4}%
\usepackage{graphicx}%
\usepackage{amsmath}%
\usepackage{amsfonts}%
\usepackage{amssymb}
\usepackage{bm}

\def\q{{ {\bm q} }}
\def\Q{{ {\bm Q} }}

\def\w{{\omega}}
\def\a{{\alpha}}

\begin{document}
\title{
Comment on ``Orbital Order, Structural Transition and Superconductivity 
in Iron Pnictides'' by Yanagi {\it et al}.}
\author{Hiroshi \textsc{Kontani}$^{1}$,
and Seiichiro \textsc{Onari}$^{2}$}
\date{\today }

\address{
$^1$ Department of Physics, Nagoya University and JST, TRIP, 
Furo-cho, Nagoya 464-8602, Japan. 
\\
$^2$ Department of Applied Physics, Nagoya University and JST, TRIP, 
Furo-cho, Nagoya 464-8602, Japan. 
}
 
\sloppy

\maketitle

Recently, Yanagi {\it et al}. in Ref. \cite{Ono} claimed that 
the development of ferro-orbital fluctuations induced by the 
``orthorhombic phonon'' is the origin of 
high-$T_{\rm c}$ superconductivity and softening of the 
shear modulus $C_{66}$ in iron pnictides.
In this comment, we explain that the orthorhombic-phonon model 
has serious difficulty in explaining high-$T_{\rm c}$.
The reason is that the orthorhombic-phonon is ``acoustic''
with the energy $\omega_\q\propto|\q|$, although the authors in Ref. \cite{Ono}
treated it as optical phonon ($\omega_{\rm o}=0.02$eV) inconsistently.

\begin{figure}[!htb]
\includegraphics[width=.9\linewidth]{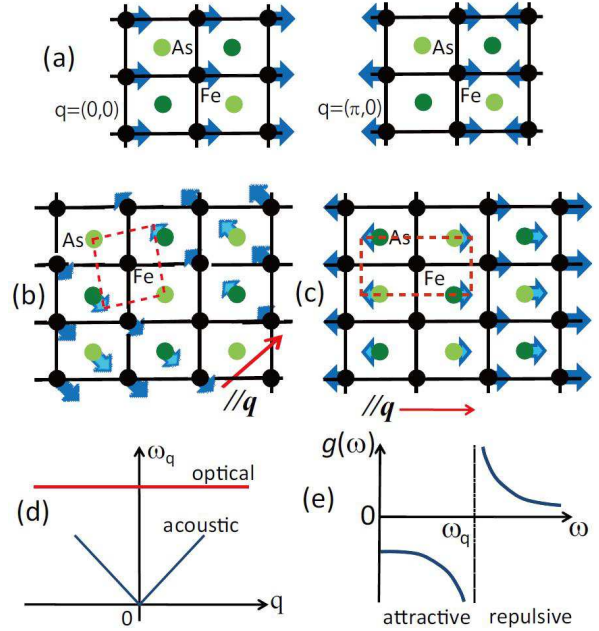}
\caption{
(a) Displacements $\bm u$ due to in-plane Fe-ion optical modes;
$\q=0$-mode and $\q=(\pi,0)$-mode.
Motions of As-ions are not shown.
(b) A transverse acoustic mode 
($\q\parallel({\bm{\hat{x}}}+{\bm{\hat{y}}})\perp{\bm u}$).
(c) A longitudinal acoustic mode
($\q\parallel{\bm u} \parallel{\bm{\hat x}}$).
(d) Energies of optical and acoustic phonons.
(e) Energy dependence of the phonon-mediated pairing interaction.
}
\label{fig:fig1}
\end{figure}

In Refs. \cite{Kontani,Saito,Onari}, we have shown that weak electron-phonon 
($e$-ph) interaction due to Fe-ion optical phonons with energy
$\w_{\rm o}\sim0.02$eV produces strong antiferro-orbital fluctuations.
For example, Fig. \ref{fig:fig1} (a) depicts 
$\q=0$ and $\q=(\pi,0)$ Fe-ion optical modes.
The derived orbital fluctuations induce the $s$-wave superconductivity
without sign reversal ($s_{++}$-wave).
A rough estimation for the transition temperature is 
\begin{eqnarray}
T_{\rm c}\sim \w_{\rm o}\exp(-1/\a\lambda),
\label{eqn:Tc}
\end{eqnarray}
where $\lambda\sim0.2$ is the dimensionless $e$-ph coupling constant,
and $\a\gg1$ is the enhancement factor for the orbital susceptibility,
$\a\sim \chi^c(\q)/\chi^0(\q)$, given by the RPA \cite{Saito}.
Since $\a\lambda\gtrsim 1$ in iron pnictides, we conclude that
high-$T_{\rm c}$ of order $\sim50$K can be realized.

In Ref. \cite{Ono}, Yanagi {\it et al}. had added an ``orthorhombic phonon''
to the model proposed in Refs. \cite{Kontani,Saito,Onari},
and claimed that the enhancement of ferro-orbital fluctuations 
due to this phonon is the origin of high-$T_{\rm c}$.
However, in contrast to the treatment in Ref. \cite{Ono},
the orthorhombic phonon is ``acoustic'' \cite{comment},
which is given by the transverse mode in Fig. \ref{fig:fig1} (b),
and is also included in the longitudinal mode in (c).
The acoustic phonon energy $\w_\q\propto |\q|$ vanishes at $\q=0$
shown in Fig. \ref{fig:fig1} (d).
Therefore, the enhancement of ferro-orbital fluctuations cannot give
high-$T_{\rm c}$ since the prefactor in eq. (\ref{eqn:Tc}) vanishes at $\q=0$.

In fact, it is hard to expect that $T_{\rm c}$ exceeds $\w_\q$ since
the phonon-mediated interaction $g_\q(\w)\propto (\w^2-\w_\q^2)^{-1}$
changes from attractive to repulsive at $\w=\w_\q$, as shown 
in Fig. \ref{fig:fig1} (e) \cite{comment3}.
Although authors in Ref. \cite{Ono} explained high-$T_{\rm c}$
by putting $\w_\q=0.02$eV even at $\q=0$ erroneously,
the orthorhombic-phonon model has serious difficulty
in explaining high-$T_{\rm c}$ in iron pnictides in reality.

Next, we discuss the softening in the shear modulus above the
orthorhombic structure transition temperature $T_S$ 
 \cite{Fernandes,Yoshizawa}.
Recently, Yoshizawa {\it et al.} \cite{Yoshizawa2}
had performed systematic measurement in Ba(Fe$_{1-x}$Co$_x$)$_2$As$_2$ for
$x=0\sim0.225$, and found that $C_{66}$ shows the largest softening 
in both under- and over-doped systems.
$C_{66}$ corresponds to the orthorhombic deformation below $T_S$,
and the structure transition ($C_{66}$-softening) is expected to 
be induced by the ferro-orbital order (fluctuations).
However, inconsistently with the claim in Ref. \cite{Ono},
the orbital order obtained in the orthorhombic-phonon model is
``incommensurate'' \cite{comment5}, which is inconsistent with 
the abovementioned experimental results.
Therefore, the origin of the orthorhombic structure transition and 
large $C_{66}$-softening is still unclear.

In summary, the orthorhombic-phonon model proposed by Ref. \cite{Ono}
has serious difficulty in explaining high-$T_{\rm c}$ in iron pnictides.
The reason is that the ``orthorhombic-phonon'' is acoustic phonon, 
although the authors treated it as optical phonon inadequately.

{\it Note added in proof}: 
Recently, the present authors found that the 
ferro-orbitalfluctuations for the $C_{66}$-softening is induced 
by the two-orbiton process with respect to the
antiferro-orbital fluctuations \cite{comment6}.



\begin{thebibliography}{99}
\bibitem{comment3}
In contrast, the sign of the spin-fluctuation propagator
$\chi_\q(\w)\propto(1+\xi^2(\q-\Q)^2-i\w/\w_{\rm sf})^{-1}$
[see SCR theory or Millis-Monien-Pines model]
does not change at $\w\sim\w_{\rm sf}$.
In case of spin-fluctuation-mediated superconductivity,
$T_{\rm c}\sim \w_{\rm sf}$ is possible.
\bibitem{comment5}
In Fig. 3(c) of Ref. \cite{Ono}, 
$\chi^c(\q,0)$ due to the orthorhombic phonon 
has an incommensurate peak with momentum $\q\approx(0.35\pi,0)$,
so the words ``ferro-orbital order'' in Fig. 4(a) 
should be corrected to ``incommensurate orbital order''.
In case of the five-orbital model in Ref. \cite{Kuroki}, 
we find that $\chi^c(\q,0)$ due to the orthorhombic phonon 
takes the maximum at $\q\approx(\pi,0)$ due to the nesting, 
while it is hardly enhanced at $\q={\bm0}$.
\end{thebibliography}
\end{document}